\documentclass[conference]{IEEEtran}
\IEEEoverridecommandlockouts
\usepackage{cite}
\usepackage{amsmath,amssymb,amsfonts}
\usepackage{algorithmic}
\usepackage{caption}
\usepackage{wrapfig}
\usepackage{subfig}
\usepackage{caption} 
\usepackage{textcomp}
\usepackage{xcolor}
\usepackage[font=footnotesize,labelfont=bf]{caption}
\usepackage{tabularx}
\usepackage{color}
\usepackage{amsmath,amsfonts,amssymb,amscd}
\usepackage[pdftex]{graphicx}
\usepackage{cite}
\usepackage{booktabs}
\usepackage{paralist}
\usepackage{enumitem}
\usepackage[export]{adjustbox}
\usepackage{multicol}
\usepackage{gensymb}
\usepackage{steinmetz}
\usepackage{mathtools}
\usepackage{siunitx}
\usepackage[draft]{hyperref}
\usepackage{wrapfig}
\usepackage{subfig}
\usepackage{caption} 
\usepackage[english]{babel}
\usepackage{blindtext}
\usepackage{verbatim}
\usepackage[compact]{titlesec}
\usepackage{secdot}

\def\BibTeX{{\rm B\kern-.05em{\sc i\kern-.025em b}\kern-.08em
    T\kern-.1667em\lower.7ex\hbox{E}\kern-.125emX}}
\begin{document}

\title{Intelligent Anomaly Mitigation in Cyber-Physical Inverter-based Systems\\
}

\author{\IEEEauthorblockN{ Asad Ali Khan, Sara Ahmed}
\IEEEauthorblockA{\textit{Department of Electrical Engineering} \\
\textit{The University of Texas at San Antonio}\\
San Antonio, United States \\
asad.khan@my.utsa.edu, sara.ahmed@utsa.edu}
\and
\IEEEauthorblockN{ Omar A. Beg}
\IEEEauthorblockA{\textit{Department of Electrical Engineering} \\
\textit{The University of Texas Permian Basin}\\
Odessa, United States \\
beg\_o@utpb.edu }

}

\maketitle

\begin{abstract}
The distributed cooperative controllers for inverter-based systems rely on communication networks that make them vulnerable to cyber anomalies. In addition, the distortion effects of such anomalies may also propagate throughout inverter-based cyber-physical systems due to the cooperative cyber layer. In this paper, an intelligent anomaly mitigation technique for such systems is presented utilizing data driven artificial intelligence tools that employ artiﬁcial neural networks. The proposed technique is implemented in secondary voltage control of distributed cooperative control-based microgrid, and results are validated by comparison with existing distributed secondary control and real-time simulations on real-time simulator OPAL-RT.

\end{abstract}


\section{Introduction}
Power electronics-based distributed generation (DG) systems are found in many applications, such as renewable energy and smart grids. Such DG systems are being adopted due to their cost-effectiveness and clean energy provision. Also, DGs improve the systems resilience and reliable operation by accommodating several energy resources with distributed control architectures.  \cite{ 9197633,9187814, 7868066, 9235913}. DGs-based microgrids extensively use inverters to produce AC power and communication-based distributed control schemes. With the inclusion of communication networks and power electronics devices, microgrids have evolved into cyber-physical systems \cite{8409327,9337864}. 
This paper considers DG systems with cooperative control that rely on communication networks, making them vulnerable to cyber anomalies. Such anomalies can compromise the control system by manipulating the sensor reading and injecting false data into measurements. The distortion effects of such anomalies in a single DG may propagate to other DGs, endangering the stability of the system by causing loss of synchronization in operation \cite{7962299,8260848,KHAN2021107024}.

An unanticipated response in the microgrid's operation is termed as an anomaly, and the cyber anomalies are initiated when an adversary targets the communication network by injecting false data or compromising the information sharing in the network\cite{7908945,9352761}.
One of the most common types of cyber anomaly is caused by false data injection (FDI) into actual measurements that may destabilize the normal operation and disrupt the communication among various agents in the network\cite{6672638}.
  \begin{figure*}[t]
\begin{minipage}{0.4\textwidth}
    \centering
    \includegraphics[width=\textwidth]{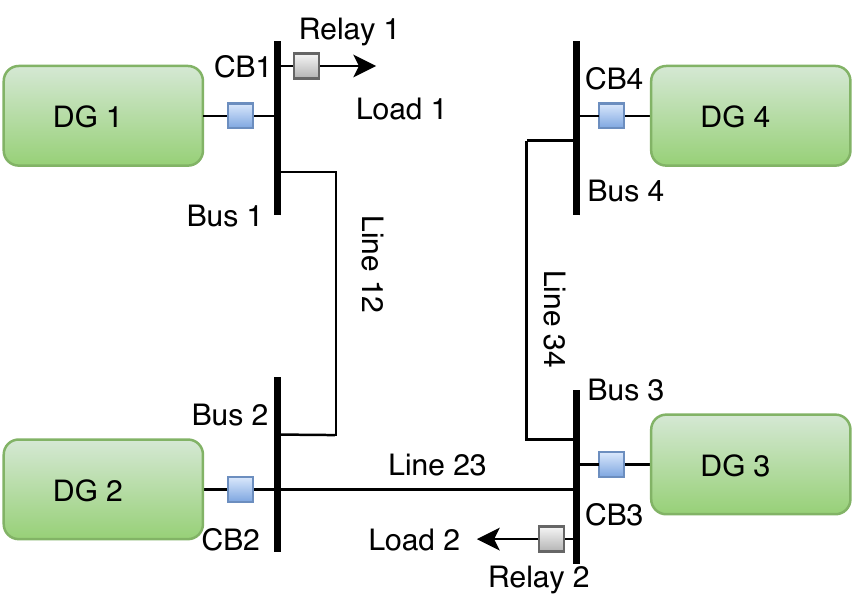}
     \caption{ \label{MG}Four DGs based  microgrid system.}
\end{minipage}
\qquad
\qquad
\qquad
\qquad
\qquad
\qquad
\begin{minipage}{0.35\textwidth}
\centering
    \includegraphics[width=\textwidth]{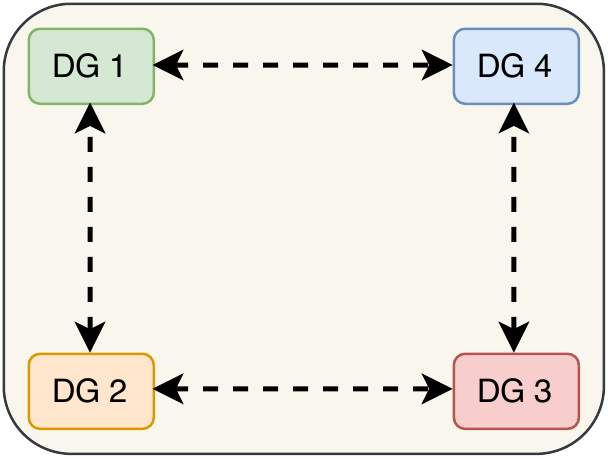}
     \caption{ \label{test11}Communication graph for the four DGs.}
 \end{minipage}
\vspace{-7pt}
\end{figure*}
Therefore, an effective mitigation framework is needed to sustain the normal operation of the microgrids.

Several learning-based intelligent control techniques for cyber anomaly identification and mitigation in microgrids exploit artificial neural networks (ANNs). An adaptive linear neuron network-based controller is used for wind turbine speed regulation and power regulation of battery energy storage system that generates the reference for electro-mechanical torque and pulse width modulation signal for the buck-boost converter of the battery system, respectively \cite{88}. 
 Learning and adaptation ability of ANN is utilized to design an adaptive control for a hybrid microgrid where a single neuron radial basis function neural network (RBFNN) is applied for maximum power point tracking of solar photovoltaic panels in \cite{89}. 
 Traditional load flow analysis algorithms such as Newton Raphson and fast decoupled pose a significant computational burden for larger power systems due to their inherent iterative nature. To overcome this, a multi-layer feed-forward neural network is used for online load flow analysis in \cite{90}. Active and reactive powers are used as input, whereas voltage magnitude and angles at various load buses are used as output for the proposed ANN training and the case studies are performed on a practical transmission network to validate the results. RBFNN based control layer is proposed in \cite{91} to address the instability issue such as large signal disturbances and undesired power sharing in hierarchical control schemes of microgrids. The proposed algorithm exploits the learning capability of ANNs using the Gaussian activation function to solve a set of power flow equations to find the optimal power sharing reference. 

 In the presence of non-linear and unbalanced loads, secondary control maintains the voltage and frequency regulation \cite{Mehmet}. Therefore, to improve the power-sharing and microgrid stability, a radial basis neural network is designed to calculate the reactive power reference in \cite{7494662}. 
 In \cite{92}, the stability of the interconnected dc distribution system is improved with a decentralized adaptive non-linear controller design that employs neural networks to mitigate voltage and power oscillations following any disturbances in the system. This neural network-based controller overcomes the unknown dynamics and stabilizes the entire grid with the help of local measurements at each converter. To overcome the difficulty in proportional-integral (PI) gains tuning and low voltage instability issues, ANN-based vector control is proposed for grid connected converters in \cite{a24}. 
 The proposed ANN controller is trained using back-propagation after introducing integral error in the input and grid voltage under disturbances in the output training data set. 
 In \cite{94}, ANN-based resilient control design is proposed along with Luenberger observer. Extended Kalman filter is used for online updation of ANN learning weights such that input to the ANN is the difference between actual system output and observed output from Luenberger observer. This way, ANN detects an anomaly in the system, and that information is used in a linear quadratic controller to compensate for the anomalies in real-time. 
 In \cite{95}, ANN-based reference tracking algorithm is introduced to mitigate the effect of FDI attacks in distributed consensus control-based DC microgrids. ANN is trained offline after running multiple non-attack cases in the proposed method, and the fine tuned ANN is used online to remove false data being injected into the communication network. 
 The estimated value of output voltage by ANN is used as a reference for secondary level control. The complexity of mathematical modeling, unavailability of a complete model of complex systems, artificial intelligence (AI) method's learning ability, adaptation to uncertain environments, and computational efficiency motivate further exploration of the AI techniques for designing resilient control.
 
 From previously mentioned research works, it is evident that  AI-based techniques are being extensively used for anomaly mitigation in microgrids. Compared to the usage of ANN as an observer in \cite{94} and in reference tracking applications for DC microgrids in \cite{95}, this work proposes the use of ANN as a resilient secondary voltage control layer. This ANN-based secondary voltage control is distributed in mechanism and implemented for AC microgrids. Distributed secondary voltage control of microgrids employs extensive communication layers for information sharing among microgrids agents. Therefore, this work focuses on the secondary layer of distributed cooperative control-based microgrids. The effectiveness of the ANN-based intelligent secondary voltage control layer is shown by various real-time case studies on real-time digital simulator OPAL-RT performed on an AC microgrid containing multiple DGs. 
 
 The rest of the paper is organized as follows. Section II has the description of the AC microgrid used in this work, along with the types of cyber anomalies. The design of ANN-based secondary voltage control is discussed in section III. In section IV, the results obtained from real-time simulations performed on the test microgrid are given. Finally, this work is concluded in section V with future directions.
\section{System Description}
 \begin{figure*}[t]
\centering
\includegraphics[width=1\textwidth]{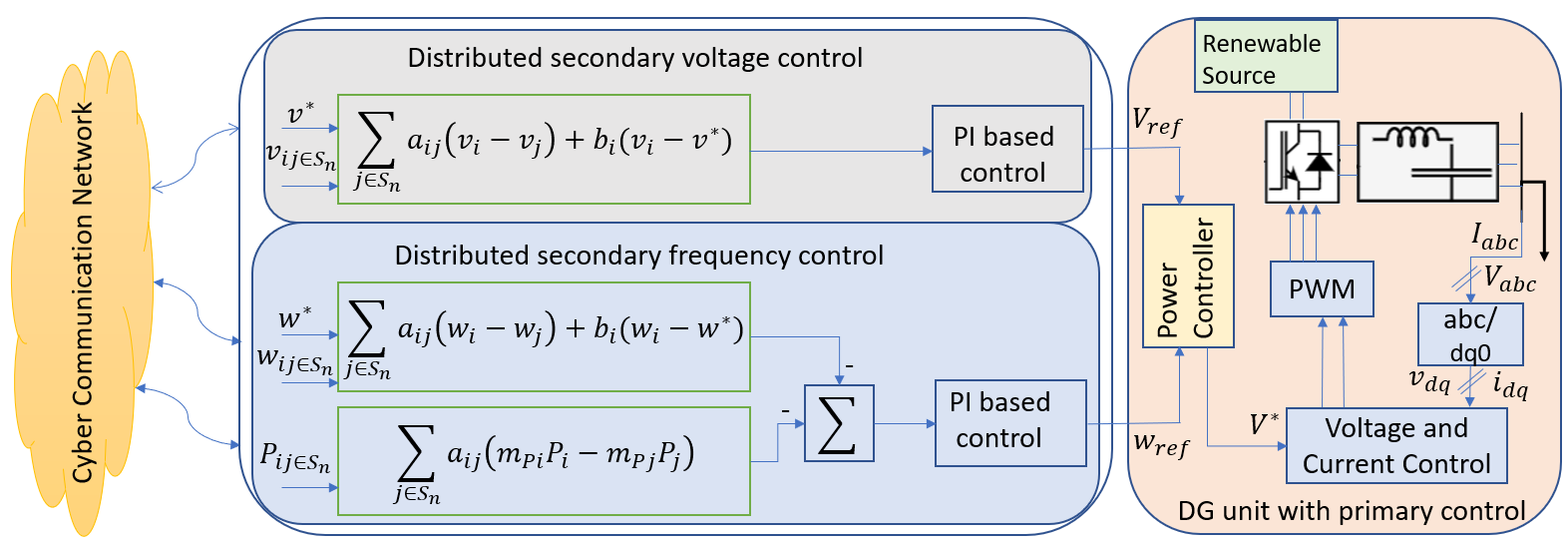}
\caption{The structure of the distributed secondary voltage and frequency control  is shown.}
\label{DS}
\end{figure*} 
\begin{figure}[t]
\centering
\includegraphics[width=.5\textwidth]{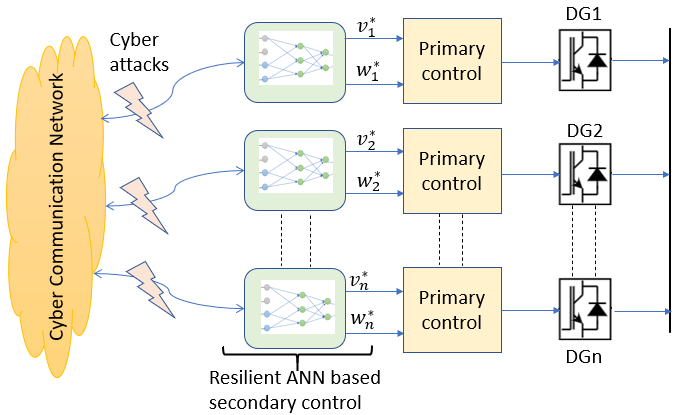}
\caption{The structure of proposed resilient ANN-based distributed secondary control  is shown.}
\label{PDS}
\end{figure} 
The microgrid system under study is composed of four DGs with distributed cooperative control as shown in Fig.~\ref{MG}. 
The distributed cooperative control is implemented at the secondary level to mitigate the voltage and frequency deviations from their nominal values which are caused by the primary control. 
The relevant control protocols are implemented over a distributed communication network as shown in Fig.~\ref{test11}. 
The secondary control selects reference for primary control such that the voltage and frequency of each DG synchronizes with their reference values, ($v^*$ and $w^*$):
\begin{equation}\label{3}
\begin{cases} 
\lim_{t\to\infty}
\left\|v_o-v^*\right\|=0,~~~\\
\lim_{t\to\infty}
\left\|w_i-w^*\right\|=0.
\end{cases}
\end{equation}
As shown in Fig. \ref{test11}, based on digraph $\mathcal{G}$, the $n^{th}$ DG, where $n \in [1,2,3,4]$, may need to share their voltage information over the communication network\cite{test}. 
\subsection{Cyber Anomalies}\label{cyber}
The cyber anomalies target the communication layer of the microgrid by either injecting false data or compromising the information sharing in the network. In this work, false data injection (FDI) is initiated by adding false data to the voltage information of neighbouring DGs that is being fed to secondary voltage control. 
The feedback signal of a controller  can be modeled as:
\begin{equation}
    h(u_i(t))=u_i(t)+\phi_i(t),
\end{equation}
where $h(u_i(t))$ is the feedback signal after false data $\phi_i(t)$ is injected by attacker in the $i^{th}$ normal feedback signal of the controller $u_i(t)$ \cite{94,95}. Two different cases of
FDI attacks by considering various  $\phi_i(t)$ are as follows: 
\begin{enumerate}
\item\textbf {Non-periodic attack:}
A non-periodic attack is initiated by adding a constant multiple $\alpha$ of the desired signal $u_i(t)$ in $h(u_i(t))$ at certain time $\tau$ during the operation of the system, given as:
\begin{equation}\label{a}
   h(u_i(t))=\left\{\begin{matrix}
u_i(t), ~~~~~~~~~~~~~~~~~~~~~if~~ t<\tau,\\ 
u_i(t) + \alpha*u_i(t),~~~~~~~if~~t>\tau.
\end{matrix}\right.
\end{equation}
\item\textbf{ Periodic attack:}
Periodic FDI attack is time varying in nature where a sinusoidal signal having time period $(\omega t)$ and amplitude $\beta$ is injected to normal signal $u_i$, given by:
\begin{equation}\label{c}
   \phi_i(t)=\left\{\begin{matrix}
0, ~~~~~~~~~~~~~~~~~~~~~~~~~~~if~~ t<\tau,\\ 
\beta sin(\omega t)*u_i(t),~~~~~~~~~if~~t>\tau.
\end{matrix}\right.
\end{equation}
\end{enumerate}

The details of resilient secondary voltage control design using ANN to withstand the malicious FDI anomalies are provided in the following section.
\section{ANN for Resilient Distributed Secondary Voltage Control Design }
\begin{figure*}[t]
  \centering
  \subfloat[Output current]{\includegraphics[width=0.34\textwidth]{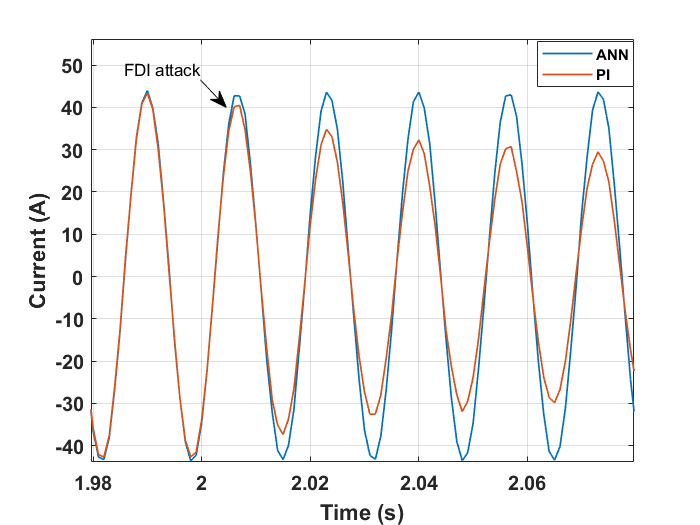}\label{I1}}
  \subfloat[Output Voltage]{\includegraphics[width=0.35\textwidth]{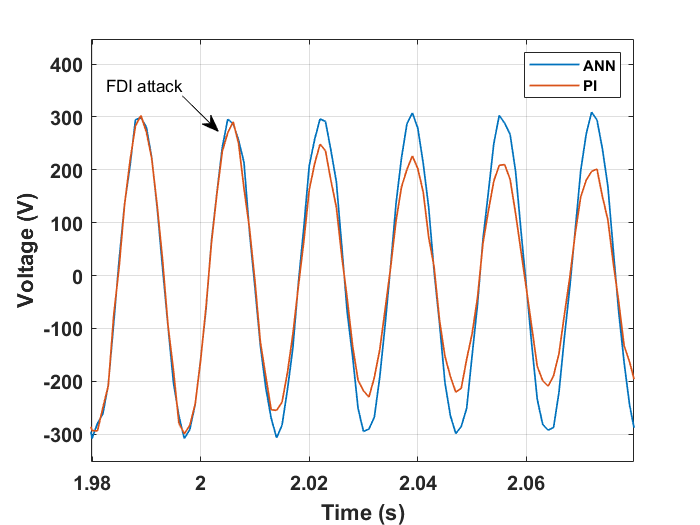}\label{V1}}
  \subfloat[Active Power]{\includegraphics[width=0.35\textwidth]{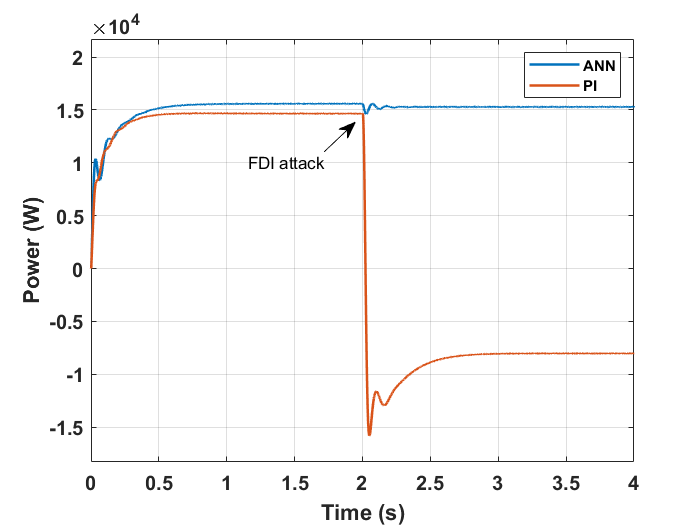}\label{P1}}
  \caption{Non-periodic FDI: The performance comparison in terms of output voltage, current at load1, and active power  at DG1 of test microgrid system is shown.}
  \label{C1V}
\end{figure*}

\begin{figure*}[t]
  \centering
  \subfloat[Output current]{\includegraphics[width=0.34\textwidth]{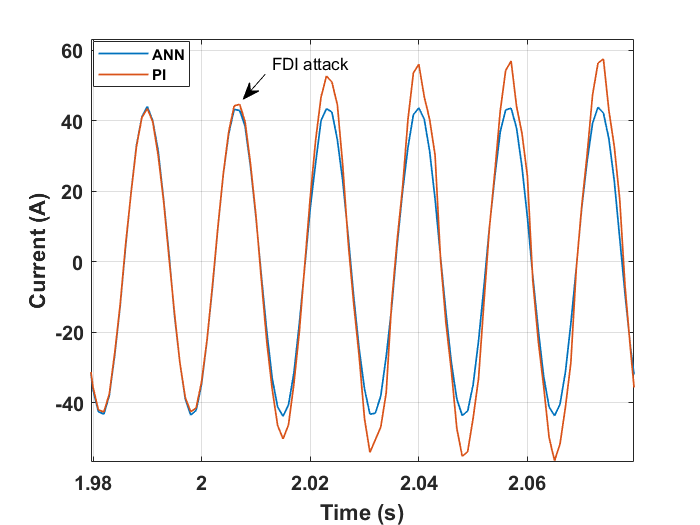}\label{I2}}
  \subfloat[Output Voltage]{\includegraphics[width=0.35\textwidth]{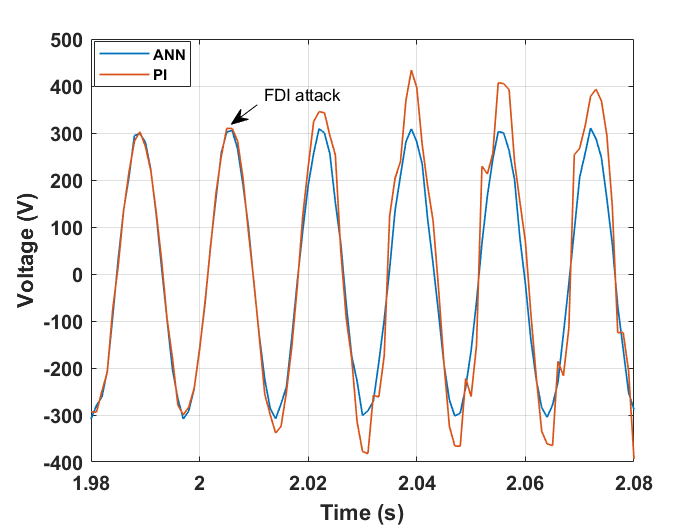}\label{V2}}
  \subfloat[Active Power]{\includegraphics[width=0.35\textwidth]{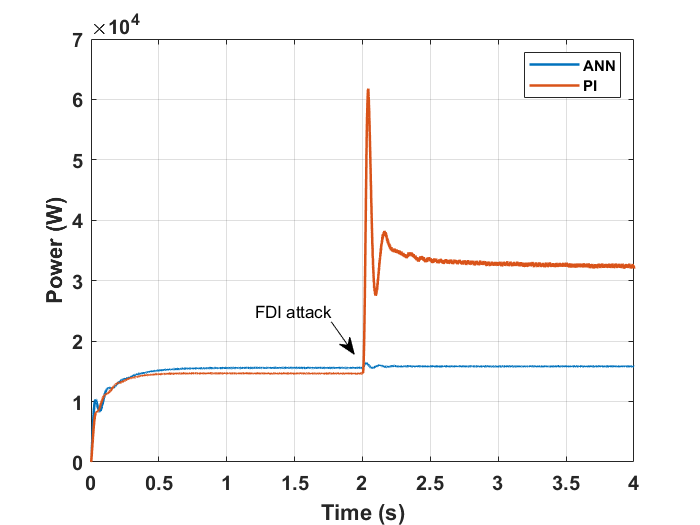}\label{P2}}
  \caption{Periodic FDI: The performance comparison in terms of output voltage, current at load1, and active power  at DG1 of test microgrid system is shown.}
  \label{C2V}
\end{figure*}
For training data generation, the secondary voltage control is selected.
This is because the secondary control layer is comprised of a  distributed cooperative control architecture with an extensive communication layer, making it vulnerable to cyber anomalies. 
The distributed secondary voltage and frequency control for each DG requires its own information and that of the neighboring DGs on the communication network, as shown in Fig. \ref{DS}.  Each DG in the microgrid shares its voltage and frequency information with the neighboring DGs as defined by the graph communication network to cooperatively implement the control objectives as shown in Fig. \ref{DS}\cite{test}. The secondary distributed cooperative control generates the reference for the primary control level implemented locally at each inverter. Therefore,  ANN-based resilient secondary voltage control is designed for each DG to generate the reference for primary control at each inverter to maintain the output voltage and current within desired permissible limits as shown in Fig. \ref{PDS}.
First, the FDI attacks are initiated at the communication links of DG1. Both time-varying and time-invariant cases are considered as given in section \ref{cyber}. While generating the training data set, the step load change is also included. The sampling time for data collection is set at 1ms.
For the complete learning of ANNs, the data is  generated for normal operating conditions and under cyber attack conditions. As shown in Fig. \ref{test11}, DG1 is sharing voltage and frequency information with DG2 and DG4 to implement the distributed cooperative control objectives. The cyber attack then target the information channels of DG1 by either jamming the communication links injecting false data into its own measurements and/or those shared by the neighboring DGs. 
\begin{figure*}[t]
  \centering
  \subfloat[Reference voltage]{\includegraphics[width=0.35\textwidth]{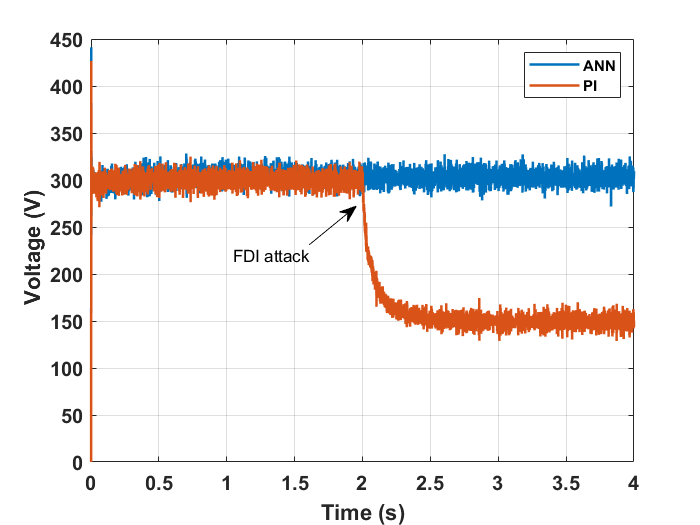}\label{vd1}}
  \subfloat[Reference frequency]{\includegraphics[width=0.35\textwidth]{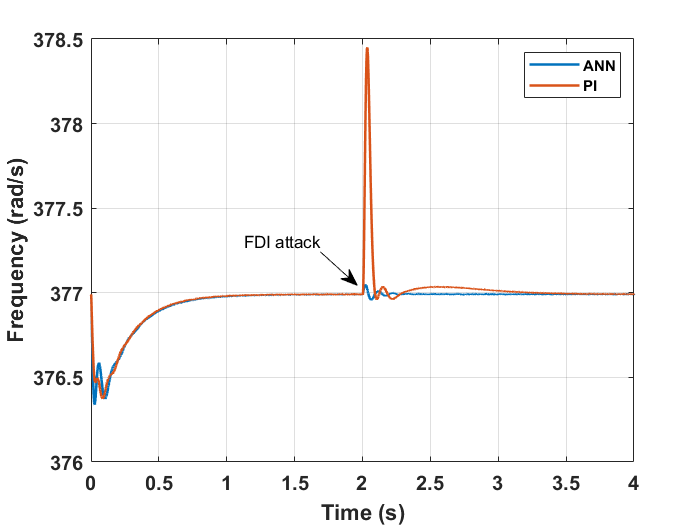}\label{f1}}
  \subfloat[Reactive Power]{\includegraphics[width=0.35\textwidth]{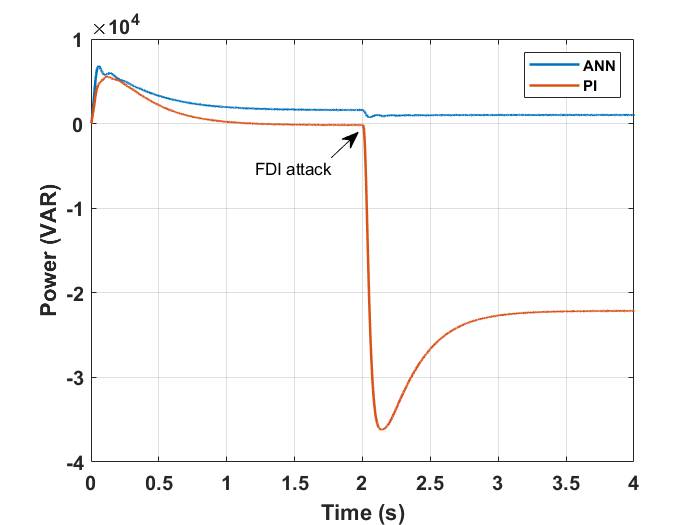}\label{Q1}}
  \caption{Non-periodic FDI: The performance comparison in terms of reference tracking and reactive power at DG1 of test microgrid system  is given.}
  \label{C1f}
\end{figure*}

\begin{figure*}[t]
  \centering
  \subfloat[Reference voltage]{\includegraphics[width=0.35\textwidth]{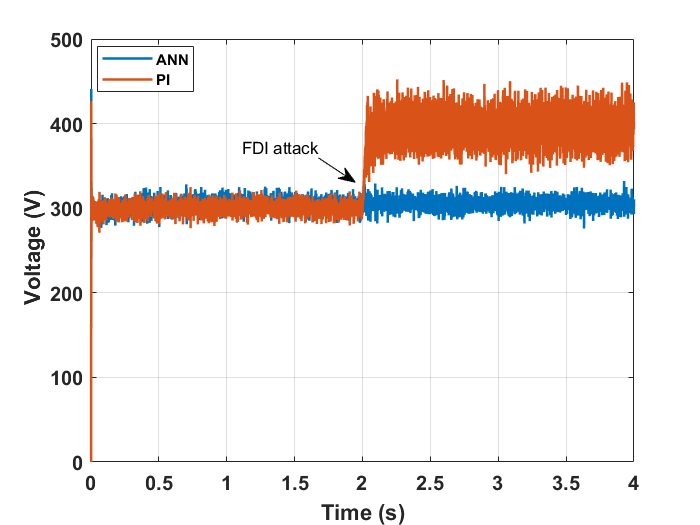}\label{vd2}}
  \subfloat[Reference frequency]{\includegraphics[width=0.35\textwidth]{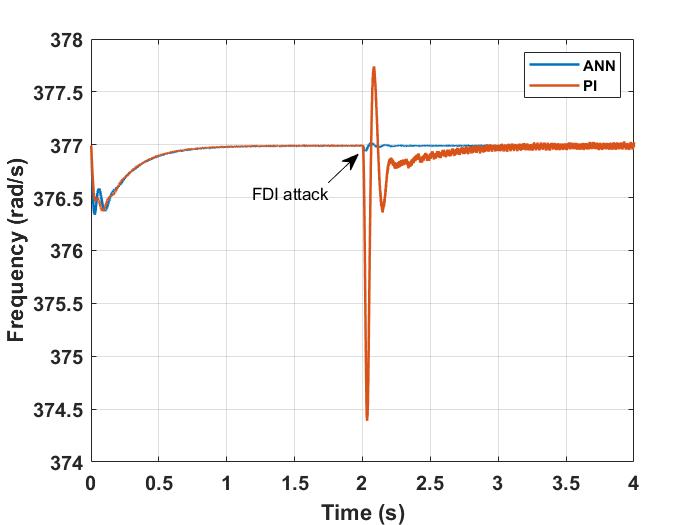}\label{f2}}
  \subfloat[Reactive Power]{\includegraphics[width=0.35\textwidth]{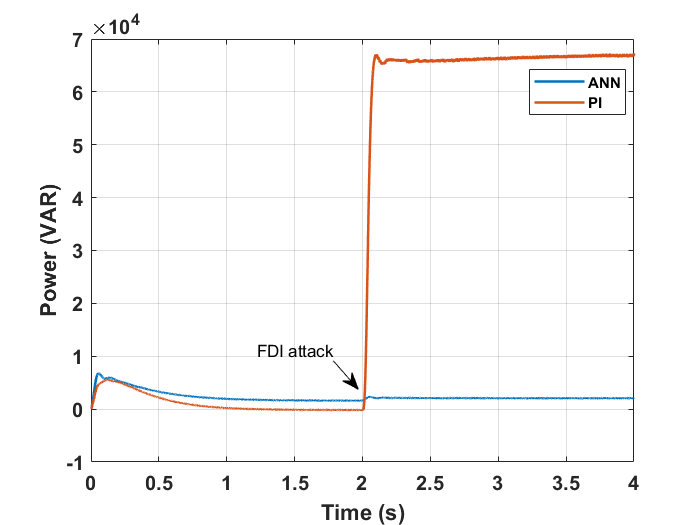}\label{Q2}}
  \caption{Periodic FDI: The performance comparison in terms of reference tracking  and reactive power at DG1 of test microgrid system  is given.}
  \label{C2f}
\end{figure*}
 The two sets of training features under normal ($x$)
and attack scenarios ($\hat{x}$) with varying load conditions are collected where, $x\in[v_{11}, v_{12}, v_{14}, v^*, v_1^*]$, and $\hat{x}\in[\hat{v}_{11}, \hat{v}_{12}, \hat{v}_{14}, \hat{v}^*, \hat{v}_1^*]$. Here $v_{ij}$ represent the voltage sharing from $DG_j$ to $DG_i$, $ij\in[1,2,3,4]$ and $v^*$ is the reference voltage.
The compromised information sharing may disturb the control objectives causing loss of synchronization or deviation from desired reference voltage value.
\textbf{Remark 1}: The FDI attack is initiated at time $t$, before the attack at $(t-\delta t)$ the output voltage follows the reference values but after the attack at $(t+\delta t)$, the output deviates from the desired reference. The difference between $\hat{v}_i^*$ and $v_i^*$ is $\left | \hat {v}_i- v_i^*\right |=\varepsilon _v,$ 
where, $\hat {v}_i$ is the output voltage of the $i^{th}$ DG under attack, ${v}_i^*$ is the reference output voltage for each DG unit, and ANN-based secondary voltage control tries to minimize this error as $\lim_{t \to \infty}\varepsilon _v=0$.
\textbf{Remark 2:} For  offline training of  ANN, the complete mathematical model of the system is not required.  The ANN learns the dynamics of the system using offline training. For the online implementation, the trained ANN model works for the system operating with the exact control mechanism used in the training phase\cite{95}. This enables the trained ANN  to perform as a distributed secondary control layer for the microgrid under study. 

The data obtained from the time series simulations of the microgrid system under various operating conditions is used as training feature vector for the resilient secondary control design as shown in Figure \ref{PDS}. 
    In case of secondary voltage control of DG1, the input feature vector is $X_i=[v_{1j};\hat {v}_{1j}; v^*]$, where $X_i \in \mathbb{R}^{1 \times 7}$ and $j\in [1,2,4]$
    , the target vector for training is $Y_i=[v_1^*]$. 
  The architecture of ANN  contains 1 input layer, 1 output layer with purelin activation function, a single hidden layer with 10 neurons having tansig activation function. 
\begin {comment}
\subsection{Training and test scenarios}
The time series simulations of the microgrid of Figre \ref{51} are performed, considering the following scenarios:
  \begin{description}
 \item [Scenario 1:] Varying load: $U_n$, where $n\in\{4,..,8\}~kW$.
 \item [Scenario 2:] Location: $ U_i$, where $i\in\{load~1, load~ 2\}$.
 \item [Scenario 3:] Cyber anomaly location: Communication links for secondary voltage and frequency control, $v_{nj}$ and $f_{nj}$, respectively, where $n\in (1,2,3,4)$, represent DG units and $j$ represent the neighboring DG units on the communication network.
  \item [Scenario 4:]Cyber anomaly type: $FDI_{nj}$, where $n\in (1,2,3,4)$, for one of the four inverters in the microgrid, and $j\in (case~1, case~2, case~3, case~4)$ shows one of the four FDI attack cases, considered in this work.
  \end{description}
    The sampling time for taking measurements is selected as 1 ms over a run time of 4 s. For cyber anomalies, three phase voltages, currents, and active power are recorded at the output of all inverters. Each inverter in the microgrid is exposed to all the four FDI and DOS attack cases as mentioned earlier. Simulations are performed for normal and anomalous conditions, with anomaly initiation at 2 s for each case. To incorporate the effect of load change, five load steps are considered for each run. This generates in total 20,000 samples for a single inverter, for all the FDI attack cases. From this sample data, voltage and frequency information  are extracted to form the feature training vector. Each case is assigned a distinct numerical target matching the desired reference value to enable the ANN control maintain the normal operation of microgrid with anomalies occurring at each inverter. 

For ANN based control, the standard rule of dividing training and testing data by the ratio 3:1 is implemented for cross validation, and to avoid overfitting. During the offline training, the ANNs are assigned unique numerical targets depending upon the desired reference value and location of the anomaly which is a common practice to train the ANNs \cite{108}.
\end{comment}
After training  the ANN, the performance of the proposed secondary voltage control is validated by running real time scenarios on real time digital simulator OPAL-RT under FDI cyber anomalies and the results are discussed in the next section.
\section{Simulation Results}
The test microgrid system, as shown in Fig. \ref{MG} is used to evaluate the performance of the proposed ANN-based secondary voltage control. The four DGs  voltage source inverters are interconnected through RL lines to supply AC power to 2 three phase RL loads, represented as $load~1$ and $load~2$ in Fig. \ref{MG}.

The secondary voltage control of DG1 is selected, and a non-periodic FDI attack is initiated at the communication link sharing the voltage measurements of DG1. This cyber anomaly is based on  \eqref{a}, where 
the attacker starts injecting the false data into the voltage measurements of  DG1 with $\alpha=0.5$, starting at $t=2$ s.
The microgrid operates in normal mode for $t<2$ s, and after that, false data is injected to compromise the performance of secondary voltage control. 
The performance of the ANN-based distributed secondary voltage control is compared to proportional integral (PI) control, in terms of output voltage and current at $load~1$ , and active power output of DG1 as shown in Figure \ref{C1V}. After initiating the periodic FDI anomaly, the proposed ANN-based secondary voltage control maintained the system within desired operating limits. In contrast, PI-based secondary voltage control did not withstand the malicious FDI and could not bring back the system to expected operating conditions. 
Also, the reference tracking capability and reactive power at the output of DG1  with ANN and PI-based secondary voltage control following a non-periodic FDI attack are shown in Figure \ref{C1f}. Again, it can be seen that ANN-based control maintained the normal operation compared to PI-based control.\\
After that, the periodic FDI anomaly based on \eqref{c} is introduced on the voltage communication link of DG1. A periodic sinusoidal signal having amplitude $\beta=0.5$ and frequency $\omega=60$ Hz is added to the voltage signal coming from DG1 at $t=2$ s. The performance of ANN and PI-based control under normal and attack scenario is shown in Fig. \ref{C2V} and Fig. \ref{C2f}. As shown in Fig. \ref{C2V}, after initiating periodic attack at $t=2$ s, the output voltage and current at $load~1$, and the active power output of DG1 deviates from their normal operating limits with the PI-based secondary control whereas ANN-based secondary voltage control sustained the normal operation of the microgrid. Similarly the reference tracking capability and reactive power output at DG1 following a periodic FDI at voltage signal of DG1 with ANN and PI-based secondary voltage control are shown in Fig. \ref{C2f}.
From these results it is evident that ANN-based distributed secondary voltage control learned the dynamics of the system efficiently and showed the resilience performance by maintaining the desired operation of test microgrid under various FDI anomalies targeting the voltage communication network.
\renewcommand*{\thesubsubsection}{\Alph{subsubsection})}

\section{Conclusion}
A new  intelligent technique for anomaly mitigation in distributed cooperative control-based microgrid  is presented. This technique utilizes the ANNs to implement the distributed secondary voltage control layer of microgrids. The ANNs are trained using time series simulations of the microgrid.
 The offline training is applied to cover the normal and attack operation of the microgrid, that may not be applied on a real system due to safety limits. 
The proposed controller depicted the improved performance by maintaining the normal operation under attack scenario compared to PI-based secondary voltage control. The proposed ANN-based secondary voltage control performance under FDI attack is validated with real-time scenarios on real-time digital simulator OPAL-RT to show the effectiveness of the proposed control. The future work will focus on expanding the ANN-based secondary control to include the frequency control to enhance the resilience of the distributed secondary control of microgrids.
\bibliographystyle{./bibliography/IEEEtran}
\bibliography{./bibliography/IEEEabrv,./bibliography/IEEEexample}

\vspace{12pt}

\end{document}